\documentclass[12pt]{article}

\usepackage{a4wide}
\usepackage[pdftex,usenames,dvipsnames]{color}
\usepackage{graphics}
\usepackage{amsfonts}

\newcommand{\nit}{\noindent}

\newcommand{\np}{\newpage}
\newcommand{\dsp}{\displaystyle}
\newcommand{\vs}[1]{\vspace{#1 ex}}
\newcommand{\hs}[1]{\hspace{#1 em}}
\newcommand{\bfr}{\begin{flushright}}
\newcommand{\efr}{\end{flushright}}
\newcommand{\bc}{\begin{center}}
\newcommand{\ec}{\end{center}}
\newcommand{\ben}{\begin{enumerate}}
\newcommand{\een}{\end{enumerate}}

\newcommand{\be}{\begin{equation}}
\newcommand{\ee}{\end{equation}}
\newcommand{\ba}{\begin{array}}
\newcommand{\ea}{\end{array}}
\newcommand{\ct}{\cite}

\newcommand{\dd}[2]{\frac{\partial{#1}}{\partial{#2}}}
\newcommand{\ag}{\alpha}
\newcommand{\bg}{\beta}
\newcommand{\gam}{\gamma}
\newcommand{\del}{\delta}

\newcommand{\ve}{\varepsilon}

\newcommand{\thg}{\theta}
\newcommand{\kg}{\kappa}
\newcommand{\lb}{\lambda}
\newcommand{\sg}{\sigma}
\newcommand{\rg}{\rho}

\newcommand{\vf}{\varphi}

\newcommand{\Gam}{\Gamma}

\newcommand{\Sg}{\Sigma}

\newcommand{\bfJ}{\bold {J}}

\newcommand{\lh}{\left(}
\newcommand{\rh}{\right)}

\newcommand{\nb}{\nabla}

\newcommand{\ctg}{\mbox{\,cotan\,}}

\begin{document}

\pagestyle{empty}
\bfr
NIKHEF/2015-001
\efr

\bc
{\Large {\bf Covariant hamiltonian spin dynamics}}\\
\vs{2}

{\Large {\bf in curved space-time}}\\
\vs{7}

{\large G.\ d'Ambrosi$^{a \ast}$, S.\ Satish Kumar$^{b \dagger}$}\\
\vs{2}

{\large J.W.\ van Holten$^{a,b \ddagger}$}
\vs{5}

\today
\ec
\vs{5}

\nit
{\footnotesize {\bf Abstract} \\
The dynamics of spinning particles in curved space-time is discussed, emphasizing the hamiltonian
formulation. Different choices of hamiltonians allow for the description of different gravitating systems. 
We give full results for the simplest case with minimal hamiltonian, constructing constants of motion
including spin. The analysis is illustrated by the example of motion in Schwarzschild space-time. 
We also discuss a non-minimal extension of the hamiltonian giving rise to a gravitational equivalent 
of the Stern-Gerlach force. We show that this extension respects a large class of known constants of 
motion for the minimal case. 
}

\vfill

\footnoterule 
\nit
{\footnotesize $^a$ Nikhef, Science Park 105, Amsterdam NL }\\
{\footnotesize $^b$ Lorentz Institute, Leiden University, Niels Bohrweg 2, Leiden NL}\\
{\footnotesize $^\ast$ {\it  gdambros@nikhef.nl}}\\
{\footnotesize $^\dagger$ {\it  satish@lorentz.leidenuniv.nl} }\\
{\footnotesize $^\ddagger$ {\it  t32@nikhef.nl} }\\

\np
\pagestyle{plain}
\pagenumbering{arabic}

\nit
{\bf \large{1 Spinning-particle dynamics}}
\vs{1}

\nit
The dynamics of angular momentum and spin of gravitating compact bodies has been a subject of great interest 
and intense investigation since the early days of relativity theory 
\cite{deSitter:1916zza,Thomas:1926dy,Frenkel:1926zz,Mathisson:1937zz,Papapetrou:1951pa,Fock:1954,Dixon:1970zza,Wald:1972sz,Hanson:1974qy,Semerak:1999qc,Semerak:2007,Plyatsko:2011gf}; for recent overviews see \cite{Schaefer:2004qh,Steinhoff:2010zz,Costa:2012cy}. 
As argued in \cite{Bailey:1975fe} there are two complementary approaches to the subject. One approach 
starts from the covariant divergence-free energy-momentum tensor of matter, which makes it possible to keep 
track of aspects of the structure of the body. The energy-momentum vector and the angular-momentum tensor 
can be constructed by computing integrals of components of the energy-momentum tensor and their first moments 
over the volume of the body, using suitable boundary conditions. Equations of motion for these quantities are then 
derived by applying the conservation law for the energy-momentum tensor of matter \cite{Mathisson:1937zz,Papapetrou:1951pa,Dixon:1970zza}.

The other approach is to construct effective equations of motion for point-like objects, which is an idealization
of a compact body, at the price of neglecting details of the internal structure by assigning the point-like
object an overall position, momentum and spin. This is also known as the spinning-particle approximation,
and is used for the semi-classical description of elementary particles as well. A large variety of models for 
spinning particles is found in the literature \cite{Frenkel:1926zz,Barut:book,Duval:1972zza,Brink:1976uf,barut:1984,Khriplovich:1996dv,Khriplovich:1998ev,Barducci:1976qu,Berezin:1976eg,Salomonson:1978sk,Rietdijk:1992tx,Barausse:2009aa}. 

In this letter we take the second point of view for the description of spinning test masses in curved space-time, 
using an effective hamiltonian formalism similar to the one introduced in ref.\ \cite{Khriplovich:1989ed}. One of the 
advantages of this description is that it can be applied to compact bodies with different types of spin dynamics, 
such as different gravimagnetic ratios. In this way specific aspects of the structure can still be accounted for.
\vs{2}

\nit
{\bf \large{2 Covariant phase-space structure}} 
\vs{1}

\nit
Hamiltonian dynamical systems are specified by three sets of ingredients: the phase space, identifying the
dynamical degrees of freedom, the Poisson-Dirac brackets defining a symplectic structure, and the hamiltonian 
generating the evolution of the system with given initial conditions by specifying a curve in the phase space 
passing through the initial point. The parametrization of phase-space is not unique, as is familiar from the
Hamilton-Jacobi theory of dynamical systems. Changes in the parametrization can be compensated 
by redefining the brackets and the hamiltonian. A convenient starting point for models with gauge-field 
interactions is the use of covariant, i.e.\ kinetic, momenta rather than canonical momenta; see \cite{vanHolten:2006xq} 
and references cited there for a general discussion, and \cite{Khriplovich:1989ed} for the application to spinning 
particles.

The spin degrees of freedom are described by an antisymmetric tensor $\Sg^{\mu\nu}$, which can be 
decomposed into two space-like four-vectors by introducing a time-like unit vector $u$: $u_{\mu} u^{\mu} = -1$, 
and defining
\be
S^{\mu} = \frac{1}{2\sqrt{-g}}\, \ve^{\mu\nu\kg\lb}\, u_{\nu} \Sg_{\kg\lb}, \hs{2}
Z^{\mu} = \Sg^{\mu\nu} u_{\nu}.
\label{2.1}
\ee
By construction both four-vectors $S$ and $Z$ are space-like:
\be
S^{\mu} u_{\mu} = 0, \hs{2} Z^{\mu} u_{\mu} = 0.
\label{2.2}
\ee
In the following we take $u$ to be the proper four-velocity of the particle. Then $S$ is the Pauli-Lubanski 
pseudo-vector, from which a magnetic dipole moment can be constructed, whilst the components of $Z$, 
which will be referred to as the Pirani vector, can be used to define an electric \ct{vanHolten:1990we} or 
mass dipole moment \cite{Costa:2011zn,Costa:2014nta}. Observe that we can invert the relations (\ref{2.1}) to write
\be
\Sg^{\mu\nu} = - \frac{1}{\sqrt{-g}}\, \ve^{\mu\nu\kg\lb}\, u_{\kg} S_{\lb} + u^{\mu} Z^{\nu} - u^{\nu} Z^{\mu}.
\label{2.3}
\ee
Therefore, if the Pirani vector vanishes: $Z = 0$ \cite{Pirani:1956tn}, the full spin tensor can be reconstructed from $S$. 
However, in non-flat space-time this is generally not the case. 

The full set of phase-space co-ordinates of a spinning particle thus consists of the position co-ordinate 
$x^{\mu}$, the covariant momentum $\pi_{\mu}$ and the spin tensor $\Sg^{\mu\nu}$, with anti-symmetric 
Dirac-Poisson brackets
\be
\ba{l}
\dsp{ \left\{ x^{\mu}, \pi_{\nu} \right\} = \del_{\nu}^{\mu}, \hs{2} 
\left\{ \pi_{\mu}, \pi_{\nu} \right\} = \frac{1}{2}\, \Sg^{\kg\lb} R_{\kg\lb\mu\nu}, }\\
 \\
\dsp{ \left\{ \Sg^{\mu\nu}, \pi_{\lb} \right\} = \Gam_{\lb\kg}^{\;\;\;\mu}\, \Sg^{\nu\kg}  
 - \Gam_{\lb\kg}^{\;\;\;\nu}\, \Sg^{\mu\kg}, }\\
 \\
\dsp{ \left\{ \Sg^{\mu\nu}, \Sg^{\kg\lb} \right\} =  g^{\mu\kg} \Sg^{\nu\lb} - g^{\mu\lb} \Sg^{\nu\kg}
 - g^{\nu\kg} \Sg^{\mu\lb} + g^{\nu\lb} \Sg^{\mu\kg}. }
\ea
\label{2.4}
\ee
The brackets imply that $\pi$ represents the generator of covariant translations, whilst the spin degrees of 
freedom $\Sg$ generate internal rotations and Lorentz transformations. It is straightforward to check that 
these brackets are closed in the sense that they satisfy the Jacobi identities for triple bracket expressions.
Thus they define a consistent symplectic structure on the phase space\footnote{We have not found this 
complete set of brackets in curved space-time in the literature. However, other sets of brackets have been 
proposed \cite{Hanson:1974qy} based on a larger set of degrees of freedom, some of which are subsequently 
removed by supplementary constraints.}. 

To get a well-defined dynamical system we need to complete the phase-space structure with a hamiltonian 
generating the proper-time evolution of the system. In principle a large variety of covariant expressions can 
be constructed; however if we impose the additional condition that the particle interacts only gravitationally 
and that in the limit of vanishing spin the motion reduces to geodesic motion, the variety is reduced to 
hamiltonians 
\be
H = H_0 + H_{\Sg},  \hs{2} H_0 = \frac{1}{2m}\, g^{\mu\nu} \pi_{\mu} \pi_{\nu},
\label{2.5}
\ee
where $H_{\Sg} = 0$ whenever $\Sg^{\mu\nu} = 0$. In this letter we focus first on the dynamics generated by 
the minimal hamiltonian $H_0$. However, we also consider an extension with \cite{Khriplovich:1996dv}
\be
H_{\Sg} = \frac{\kg}{4}\, R_{\mu\nu\kg\lb} \Sg^{\mu\nu} \Sg^{\kg\lb},
\label{2.6}
\ee
The choice of hamiltonians can be enlarged further by including charges coupling the particle to vector fields 
like the electromagnetic field \cite{Khriplovich:1989ed,vanHolten:1990we}. 
\vs{2}

\np
\nit
{\bf {\large 3 Equations of motion}}
\vs{1}

\nit
Eqs.\ (\ref{2.4}) and (\ref{2.5}) specify a complete and consistent dynamical scheme for spinning particles. 
Note that the choice of hamiltonian is fixed by further physical requirements, and can differ for different compact 
objects. In that sense the hamiltonian is an {\em effective} hamiltonian, suitable to describe the motion of various
types of objects in so far as the role of other internal degrees of freedom can be restricted to their effects on overall 
position, linear momentum and spin. 

The simplest model is obtained by restricting the hamiltonian to the minimal geodesic term $H_0$. By itself 
this hamiltonian generates the following set of proper-time evolution equations:
\be
\dot{x}^{\mu} = \left\{ x^{\mu}, H_0 \right\} \hs{1} \Rightarrow \hs{1} \pi_{\mu} = m g_{\mu\nu} \dot{x}^{\nu}, 
\label{3.1}
\ee
stating that the covariant momentum $\pi$ is a tangent vector to the world line, proportional to the proper 
four-velocity $u = \dot{x}$. Next
\be
\dot{\pi}_{\mu} = \left\{ \pi_{\mu}, H_0 \right\} \hs{1} \Rightarrow \hs{1} 
 D_{\tau} \pi_{\mu} \equiv \dot{\pi}_{\mu} - \dot{x}^{\lb} \Gam_{\lb\mu}^{\;\;\;\nu} \pi_{\nu} 
 = \frac{1}{2m}\, \Sg^{\kg\lb} R_{\kg\lb\mu}^{\;\;\;\;\;\,\nu}\, \pi_{\nu},
\label{3.2}
\ee
which specifies how the world line curves in terms of the evolution of its tangent vector.
Finally the rate of change of the spin tensor is
\be
\dot{\Sg}^{\mu\nu} = \left\{ \Sg^{\mu\nu}, H_0 \right\} \hs{1} \Rightarrow \hs{1}
D_{\tau} \Sg^{\mu\nu} \equiv \dot{\Sg}^{\mu\nu} + \dot{x}^{\lb} \Gam_{\lb\kg}^{\;\;\;\mu} \Sg^{\kg\nu} 
 + \dot{x}^{\lb} \Gam_{\lb\kg}^{\;\;\;\nu} \Sg^{\mu\kg} = 0.
\label{3.3}
\ee
In these equations the overdot denotes an ordinary derivative w.r.t.\ proper time $\tau$, whereas $D_{\tau}$ 
denotes the pull-back of the covariant derivative along the world line $x^{\mu}(\tau)$. By substitution of
eq.\ (\ref{3.1}) into eq.\ (\ref{3.2}) one finds that
\be
D_{\tau}^2 x^{\mu} = \ddot{x}^{\mu} + \Gam_{\lb\nu}^{\;\;\;\,\mu} \dot{x}^{\lb} \dot{x}^{\nu} 
 =  \frac{1}{2m}\, \Sg^{\kg\lb} R_{\kg\lb\;\,\nu}^{\;\;\;\,\mu}\, \dot{x}^{\nu},
\label{3.2.a}
\ee
which reduces to the geodesic equation in the limit $\Sg = 0$. The world line is the solution of the combined 
equations (\ref{3.2.a}) and (\ref{3.3}) satisfying some initial conditions. This world line is a curve in space-time 
along which the spin tensor is covariantly constant. It has been remarked by many authors 
\cite{Costa:2012cy,Khriplovich:1989ed,Khriplovich:2008ni,vanHolten:1992hs}, that the 
spin-dependent force (\ref{3.2}) exerted by the space-time curvature on the particle is similar to the Lorentz
force with spin replacing the electric charge and curvature replacing the electromagnetic field strength. In 
this analogy the covariant conservation of spin along the world line is the natural equivalent of the conservation 
of charge. 

Even though the spin tensor is covariantly constant, this does not hold for the Pauli-Lubanski and Pirani vectors $S$ 
and $Z$ individually. Indeed, due to the gravitational Lorentz force
\be
\ba{l}
\dsp{ D_{\tau} S^{\mu} = \frac{1}{4m\sqrt{-g}}\, \ve^{\mu\nu\kg\lb} \Sg_{\kg\lb} \Sg^{\ag\bg} R_{\ag\bg\nu\rg} u^{\rg}, }\\
 \\
\dsp{ D_{\tau} Z^{\mu} = \frac{1}{2m}\, \Sg^{\mu\nu} \Sg^{\ag\bg} R_{\ag\bg\nu\rg} u^{\rg}, }
\ea
\label{3.4}
\ee
where $\Sg^{\mu\nu}$ is the linear expression in terms of $S^{\mu}$ and $Z^{\mu}$ given in eq.\ (\ref{2.3}). 
We observe that the rate of change of both spin vectors is of order ${\cal O}[\Sg^2]$. In particular, as $Z$ is 
not conserved in non-flat space-times the condition $Z = 0$ cannot be imposed during the complete motion 
in general. Indeed, the evolution of the system is completely determined by eqs.\ (\ref{3.1}, \ref{3.2}, \ref{3.3}), 
and leaves no room for additional constraints. 

We close this section by remarking that the gravitational Lorentz force for unit mass
$1/2\, \Sg^{\kg\lb} R_{\kg\lb\;\,\nu}^{\;\;\;\,\mu} u^{\nu}$ can be interpreted geometrically as the change 
in the unit vector $u^{\mu}$ generated by transporting it around a closed loop with area projection in the 
$x^{\kg}$-$x^{\lb}$-plane equal to $\Sg^{\kg\lb}$. 
\vs{2}

\nit
{\bf \large{4 Conservation laws}}
\vs{1}

\nit
By construction the time-independent hamiltonian represented by (\ref{2.5}), (\ref{2.6}) is a constant of motion
for the spinning body, irrespective of the specific geometry of the space-time manifold. In particular for the 
minimal geodesic hamiltonian $H_0$ we have
\be
H_0 = - \frac{m}{2}.
\label{4.0}
\ee
Another obvious constant of motion is the total spin:
\be
I =  \frac{1}{2}\, g_{\kg\mu} g_{\lb\nu} \Sg^{\kg\lb} \Sg^{\mu\nu} = S_{\mu} S^{\mu} + Z_{\mu} Z^{\mu}.
\label{4.0.0}
\ee
In addition, there may exist conserved quantities $J(x,\pi,\Sg)$ resulting from symmetries of the background 
geometry, as implied by Noether's theorem \cite{Dixon:1970zza,Semerak:1999qc,Rudiger:1981}. They are solutions of the generic equation 
\be 
\left\{ J, H_0 \right\} = \frac{1}{m}\, g^{\mu\nu} \pi_{\nu} \left[ \dd{J}{x^{\mu}} 
 + \Gam_{\mu\lb}^{\;\;\;\kg}\, \pi_{\kg} \dd{J}{\pi_{\lb}} + \frac{1}{2}\, \Sg^{\ag\bg} R_{\ag\bg\lb\mu} \dd{J}{\pi_{\lb}} 
 + \Gam_{\mu\ag}^{\;\;\;\kg}\,  \Sg^{\lb\ag} \dd{J}{\Sg^{\kg\lb}} \right] = 0.
\label{4.1}
\ee
It follows that any constants of motion linear in momentum \cite{Rudiger:1981} are of the form
\be
J = \ag^{\mu} \pi_{\mu} + \frac{1}{2}\, \bg_{\mu\nu}\, \Sg^{\mu\nu}, 
\label{4.2}
\ee
with
\be
\nb_{\mu} \ag_{\nu} + \nb_{\nu} \ag_{\mu} = 0, \hs{2} \nb_{\lb} \bg_{\mu\nu} = R_{\mu\nu\lb}^{\;\;\;\;\;\,\kg} \ag_{\kg}.
\label{4.3}
\ee
These equations imply that $\ag$ is a Killing vector on the space-time, and $\bg$ is its anti-symmetrized gradient:
\be
\bg_{\mu\nu} = \frac{1}{2} \lh \nb_{\mu} \ag_{\nu} - \nb_{\nu} \ag_{\mu} \rh.
\label{4.4}
\ee
Similarly constants of motion quadratic in momentum \cite{Rudiger:1983} are of the form:
\be
J = \frac{1}{2}\, \ag^{\mu\nu} \pi_{\mu} \pi_{\nu} + \frac{1}{2} \bg_{\mu\nu}^{\;\;\;\lb}\, \Sg^{\mu\nu} \pi_{\lb} 
 + \frac{1}{8}\, \gam_{\mu\nu\kg\lb} \Sg^{\mu\nu} \Sg^{\kg\lb},
\label{4.5}
\ee
where the coefficients have to satisfy the ordinary partial differential equations
\be
\ba{l}
\nb_{\lb} \ag_{\mu\nu} + \nb_{\mu} \ag_{\nu\lb} + \nb_{\nu} \ag_{\lb\mu} = 0, \\
 \\
\nb_{\mu} \bg_{\kg\lb\nu} + \nb_{\nu} \bg_{\kg\lb\mu} = R_{\kg\lb\mu}^{\;\;\;\;\;\,\rg} \ag_{\nu\rg} + R_{\kg\lb\nu}^{\;\;\;\;\;\,\rg} \ag_{\mu\rg}, \\
 \\
\nb_{\rg} \gam_{\mu\nu\kg\lb} = R_{\mu\nu\rg}^{\;\;\;\;\;\,\sg} \bg_{\kg\lb\sg} + R_{\kg\lb\rg}^{\;\;\;\;\;\,\sg} \bg_{\mu\nu\sg}.
\ea
\label{4.6}
\ee
Thus $\ag$ is a symmetric rank-two Killing tensor, and the coefficients $(\bg,\gam)$ satisfy a hierarchy of 
inhomogeneous Killing-like equations determined by the $\ag_{\mu\nu}$. In the case of Grassmann-valued spin 
tensors $\Sg^{\mu\nu} = i \psi^{\mu} \psi^{\nu}$ the coefficient $\gam$ is completely anti-symmetric and the 
equations are known to have a solution  in terms of Killing-Yano tensors \cite{Gibbons:1993ap}. 
 
The constants of motion (\ref{4.2}) linear in momentum are special in that they define a Lie algebra: if $J$ and 
$J'$ are two such constants of motion, then their bracket is a constant of motion of the same type. This follows 
from the Jacobi identity
\be
\left\{ \left\{ J, J' \right\}, H_0 \right\} = \left\{ \left\{ J, H_0 \right\}, J' \right\} - \left\{ \left\{ J', H_0 \right\}, J \right\}  = 0.
\label{4.7}
\ee
Thus, if $\{ e_i \}_{i=1}^r$ is a complete basis for Killing vectors:
\[
\ag^{\mu} = \ag^i e_i^{\mu}, \hs{2} e_j^{\nu} \nb_{\nu} e_i^{\mu} - e_i^{\nu} \nb_{\nu} e_j^{\mu} = f_{ij}^{\;\;k} e_k^{\mu},
\]
the constants of motion define a representation of the same algebra: 
\be
J_i = e_i^{\mu} \pi_{\mu} + \frac{1}{2}\, \nb_{\mu} e_{i\nu}\, \Sg^{\mu\nu} \hs{1} \Rightarrow \hs{1}
 \left\{ J_i, J_j \right\} = f_{ij}^{\;\;k} J_k.
\label{L2.5}
\ee
Evidently such constants of motion are helpful in the analysis of spinning particle dynamics \cite{Semerak:1999qc,Plyatsko:2011gf,VanHolten:1997ur}. 
\vs{2}

\nit
{\bf \large{5 Schwarzschild space-time}}
\vs{1}

\nit
The dynamics of spinning bodies can be illustrated by the motion in a static and spherically symmetric 
Schwarzschild space-time, for which the hamiltonian $H_0$ in Droste co-ordinates is given by
\be
2 m H_0 = - \frac{1}{1 - \frac{2M}{r}}\, \pi_t^2 + \lh 1 - \frac{2M}{r} \rh \pi_r^2 + r^2 \pi_{\thg}^2 + 
 r^2 \sin^2 \thg\, \pi_{\vf}^2.
\label{L3.1}
\ee 
The space-time manifold admits four Killing vectors, for time-translations and rotations. They give rise 
to the conservation of kinetic energy:
 \be
 - E = \pi_t + \frac{M}{r^2}\, \Sg^{tr}, 
 \label{L3.2}
 \ee
 and angular momentum: 
 \be
 \ba{lll}
J_1 & = & \dsp{ - \sin \vf\, \pi_{\thg} - \ctg \thg \cos \vf\, \pi_{\vf} }\\
 & & \\
 & & \dsp{ - r \sin \vf\, \Sg^{r\thg} - r \sin \thg \cos \thg \cos \vf\, \Sg^{r\vf} + r^2 \sin^2 \thg \cos \vf\, \Sg^{\thg\vf}, }\\
 & & \\
J_2 & = & \dsp{ \cos \vf\, \pi_{\thg} - \ctg \thg \sin \vf\, \pi_{\vf} }\\
 & & \\
 & & \dsp{ + r \cos \vf\, \Sg^{r\thg} - r \sin \thg \cos \thg \sin \vf\, \Sg^{r\vf} + r^2 \sin^2 \thg \sin \vf\, \Sg^{\thg\vf}, }\\
 & & \\
J_3 & = & \dsp{ \pi_{\vf} + r \sin^2 \thg\, \Sg^{r\vf} + r^2 \sin \thg \cos \thg\, \Sg^{\thg\vf}. }
\ea 
 \label{L3.3}
 \ee
It is straightforward to check that these satisfy the usual algebra of time-translations and spatial rotations:
\be
\left\{ E, J_i \right\} = 0, \hs{2} \left\{ J_i, J_j \right\} = \ve_{ijk} J_k.
\label{L3.4}
\ee
As usual, the conservation of total angular momentum and the spherical symmetry of the space-time geometry 
allow one to take the angular momentum $\bfJ$ as the direction of the $z$-axis, such that 
\be
\bfJ = (0, 0, J).
\label{L15.1}
\ee
For spinless particles, for which the angular momentum is strictly orbital, this implies that the orbital motion is 
in a plane perpendicular to the angular momentum 3-vector; with our choice of the $z$-axis this is the equatorial 
plane $\thg = \pi/2$.

In the presence of spin the result no longer holds in general, as the precession of spin can be compensated by precession
of the orbital angular momentum, resulting in a non-planar orbit \cite{Bini:2005nt}. However, one can ask under which conditions
planar motion is still possible. As in that case the directions of orbital and spin angular momentum are separately preserved,
it means that necessary conditions for motion in the equatorial plane are 
\be
J_1 = J_2 = 0, \hs{2} \pi_{\thg} = 0,
\label{L15.2}
\ee 
and therefore also
\be
\Sg^{r\thg} = \Sg^{\thg\vf} = 0.
\label{L15.3}
\ee
Furthermore the absence of acceleration perpendicular to the equatorial plane expressed by $D_{\tau} \pi_{\thg} = 0$
implies that
\be
\Sg^{t\thg} = 0.
\label{L15.4}
\ee
Thus planar motion requires alignment of the spin with the orbital angular momentum; it is straighforward to show
that the reverse statement also holds \cite{Bini:2014poa,d'Ambrosi:preparation}. 

In terms of the four-velocity components we are now left with relevant constants of motion
\be
E = m \lh 1 - \frac{2M}{r} \rh u^t  - \frac{M}{r^2}\, \Sg^{tr}, 
\label{L3.5}
\ee
and
\be
J  =  m r^2 u^{\vf} + r \Sg^{r\vf},
\label{L3.6}
\ee
in addition to the hamiltonian constraint
\be
\lh 1 - \frac{2M}{r} \rh u^{t\,2} = 1 + \frac{u^{r\,2}}{1 - \frac{2M}{r}} + r^2 u^{\vf\,2},
\label{L15.5}
\ee
and the conservation of total spin $I$, or equivalently:
\be
\Sg^{t\vf\,2} = - \frac{1}{r^2} \frac{I + \Sg^{tr\,2}}{1 - \frac{2M}{r}} + \frac{\Sg^{r\vf\,2}}{\lh 1 - \frac{2M}{r} \rh^2}.
\label{L15.6}
\ee
These equations show, that once the orbital velocities are known, all the non-vanishing spin components 
can be calculated from eqs.\ (\ref{L3.5}), (\ref{L3.6}) and (\ref{L15.6}).

The simplest type of planar orbit is the circular orbit $r = R$ = constant, $u^r = 0$. In this case the symmetry of
the orbit implies that $(u^t, u^{\vf})$ are constant in time, and that $\Sg^{t\vf} = 0$. This can be shown as follows.
First, absence of radial acceleration $D_{\tau} u^r = 0$ gives, upon using the conservation laws for $E$ and $J$:
\be
\lh 1 - \frac{2M}{R} \rh \lh 2 - \frac{3M}{R} \rh m u^{t\,2} - \lh 1 - \frac{3M}{R} \rh m R^2 u^{\vf\,2} = 
 2E \lh 1 - \frac{2M}{R} \rh u^t + \frac{JM}{R}\, u^{\vf},
\label{L15.7}
\ee
whilst the hamiltonian constraint (\ref{L15.5}) simplifies to
\be
\lh 1 - \frac{2M}{R} \rh u^{t\,2} = 1 + R^2 u^{\vf\,2}.
\label{L15.8}
\ee
These two equations can be solved for $u^t$ and $u^{\vf}$ in terms of $(R, E, J)$, implying that they are constant. 
An immediate consequence is, that $\Sg^{tr}$, $\Sg^{r\vf}$ and $\Sg^{t\vf}$ are constant as well, and actually
$\Sg^{t\vf}$ vanishes. This follows directly from the absence of four-acceleration:
\be
\frac{du^t}{d\tau} = \frac{M}{mR}\, u^{\vf} \Sg^{t\vf} = 0, \hs{2}
\frac{du^{\vf}}{d\tau} = \frac{M}{mR^3} \lh 1 - \frac{2M}{R} \rh u^t \Sg^{t\vf} = 0.
\label{L15.9}
\ee
Then also the rate of change of $\Sg^{t\vf}$ must vanish:
\be
- \frac{M}{R} \lh 1 - \frac{2M}{R} \rh \frac{d\Sg^{t\vf}}{d\tau} = \lh 1 - \frac{M}{R} \rh \lh 1 - \frac{3M}{R} \rh m u^t u^{\vf} 
 + \frac{JM^2}{R^4}\, u^t - E \lh 1 - \frac{2M}{R} \rh u^{\vf} = 0. 
\label{L15.9.1}
\ee
Now from eqs.\ (\ref{L15.7}) and (\ref{L15.8}) it follows that 
\be
\frac{2E}{m} \lh 1 - \frac{2M}{R} \rh u^t = 2 - \frac{3M}{R} - \frac{JM}{mR}\, u^{\vf} + R^2 u^{\vf\,2}.
\label{L15.11}
\ee
These equations then allow the elimination of $E$ and $u^t$, with the result that
\be
\frac{JM}{mR^2} \lh \frac{2M}{R} + R^2 u^{\vf\,2} \rh = Ru^{\vf} \left[ \frac{M}{R} - 
 \lh 1 - \frac{6M}{R} + \frac{6M^2}{R^2} \rh R^2 u^{\vf\,2} \right].
\label{L15.12}
\ee
As for the total spin, for circular orbits the expression (\ref{L15.6}) can be written as
\be
\ba{lll}
I & = & \dsp{  - \Sg^{tr\,2} + \frac{R^2 \Sg^{r\vf\,2}}{1 - \frac{2M}{R}} }\\
 & & \\
 & = & \dsp{ - \frac{R^4}{M^2} \left[ \lh 1 - \frac{2M}{R} \rh m u^t - E \right]^2 
   + \frac{1}{\lh 1 - \frac{2M}{R} \rh^2} \left[ J - m R^2 u^{\vf} \right]^2. }
\ea
\label{L15.10} 
\ee
Thus for circular orbits $u^{\vf}$ and $u^t$ are constants which can be expressed in terms of $R$ 
and $J$, in turn fixing $E$ and $I$ as well.
\vs{2}

\nit
{\bf \large{6 Non-minimal hamiltonians}}
\vs{1}

\nit
So far we have studied the dynamics of compact spinning objects generated by the minimal geodesic hamiltonian 
$H_0$. In this section we consider the non-minimal extension (\ref{2.6})
\[
H_{\Sg} = \frac{\kg}{4}\, R_{\mu\nu\kg\lb} \Sg^{\mu\nu} \Sg^{\kg\lb},
\] 
including the spin-spin interaction via space-time curvature.  It is straightforward to derive the equations of motion:
\be
\ba{ll}
\dot{x}^{\mu} = \left\{ x^{\mu}, H \right\} & \Rightarrow \hs{1} \pi_{\mu} = m g_{\mu\nu} \dot{x}^{\nu}, \\
 & \\
\dsp{ \dot{\pi}_{\mu} = \left\{ \pi_{\mu}, H \right\} }& \dsp{ \Rightarrow \hs{1} 
 D_{\tau} \pi_{\mu} = \frac{1}{2m}\, \Sg^{\kg\lb} R_{\kg\lb\mu}^{\;\;\;\;\;\,\nu} \pi_{\nu} - \frac{\kg}{4}\, \Sg^{\kg\lb}
 \Sg^{\rg\sg} \nb_{\mu} R_{\kg\lb\rg\sg}, }\\
 & \\
\dsp{ \dot{\Sg}^{\mu\nu} = \left\{ \Sg^{\mu\nu}, H \right\} }& \dsp{ \Rightarrow \hs{1} D_{\tau} \Sg^{\mu\nu} = 
  \kg \Sg^{\kg\lb} \lh R_{\kg\lb\;\,\sg}^{\;\;\;\,\mu} \Sg^{\nu\sg} - R_{\kg\lb\;\,\sg}^{\;\;\;\,\nu} \Sg^{\mu\sg} \rh.  } 
\ea
\label{6.1}
\ee
Comparing again with the electro-magnetic force, the middle equation implies that in addition to the gravitational 
Lorentz force there is a gravitational Stern-Gerlach force, coupling spin to the gradient of the curvature. Therefore
the coupling parameter $\kg$ has been termed the gravimagnetic ratio \cite{Khriplovich:1998ev,Khriplovich:1997ni}. Like in the electromagnetic case \cite{Heinemann:1996aq} the Pauli-Lubanski and Pirani-vectors are affected by this Stern-Gerlach force:
\be
\ba{l}
\dsp{ D_{\tau} S^{\mu} = \frac{1}{4m\sqrt{-g}}\, \ve^{\mu\nu\kg\lb} \Sg_{\kg\lb} \Sg^{\ag\bg} \lh R_{\ag\bg\nu\sg} u^{\sg} 
 - \frac{\kg}{2}\, \Sg^{\rg\sg} \nb_{\nu} R_{\rg\sg\ag\bg} \rh, }\\
 \\
\dsp{ D_{\tau} Z^{\mu} = - \kg \Sg^{\kg\lb} R_{\kg\lb\;\,\nu}^{\;\;\;\,\mu} Z^{\nu} 
 + \lh \kg + \frac{1}{2m} \rh \Sg^{\mu\nu} \Sg^{\kg\lb} R_{\kg\lb\nu\sg} u^{\sg} 
 - \frac{\kg}{4m}\, \Sg^{\mu\nu} \Sg^{\kg\lb} \Sg^{\rg\sg} \nb_{\nu} R_{\kg\lb\rg\sg}. }
\ea
\label{6.2}
\ee
The second equation simplifies strongly for the special value
\be
\kg = - \frac{1}{2m}.
\label{6.3}
\ee
In that case an initial condition $Z^{\mu} = 0$ is conserved up to terms of cubic order in spin. 

For the extended hamiltonian the conditions for the existence of constants of motion are modified.
The total spin $I$ defined in (\ref{4.0.0}) is still conserved, but the conserved hamiltonian now is of 
course $H = H_0 + H_{\Sg}$. Finally we prove that the constants of motion $J$ of the form (\ref{4.2}) 
are preserved under this modification of the hamiltonian. To see this, observe that
\be
\left\{ J, H_{\Sg} \right\} = - \kg \Sg^{\mu\nu} \Sg^{\rg\sg} \lh \frac{1}{4}\, \ag^{\lb} \nb_{\lb} R_{\mu\nu\rg\sg}
 + \bg_{\mu\lb} R^{\lb}_{\;\,\nu\rg\sg} \rh.
\label{6.4}
\ee
For the Killing-vector solutions (\ref{4.3}) the right-hand side takes the form
\be
\ba{lll}
\dsp{ \Sg^{\mu\nu} \Sg^{\rg\sg} \lh \frac{1}{4}\, \ag^{\lb} \nb_{\lb} R_{\mu\nu\rg\sg} 
 + \bg_{\mu\lb} R^{\lb}_{\;\,\nu\rg\sg} \rh }& = & \dsp{ \frac{1}{2}\, \Sg^{\mu\nu} \Sg^{\rg\sg}  
  \lh \nb_{\mu} \nb_{\rg} \nb_{\sg} + \nb_{\rg} \nb_{\mu} \nb_{\sg} \rh  \ag_{\nu} }\\
 & & \\
 & = & \dsp{ \frac{1}{2}\, \Sg^{\mu\nu} \Sg^{\rg\sg} \lh \nb_{\mu} \nb_{\rg} + \nb_{\rg} \nb_{\mu} \rh \bg_{\sg\nu}
 = 0, }
\ea
\label{6.5}
\ee
due to the anti-symmetry of the tensor $\bg_{\sg\nu}$. Therefore in particular the expressions (\ref{L3.2}) 
and (\ref{L3.3}) also define constants of motion in Schwarzschild space-time in the presence of Stern-Gerlach
forces, as described by the non-minimal hamiltonian (\ref{2.6}).
\vs{2}

\nit
{\bf \large{7 Conclusions}}
\vs{1}

\nit
In the context of general relativity the notion of point-masses is troublesome; any non-zero mass has a characteristic 
scale, typified by its Schwarzschild radius, describing its minimal size as defined by the corresponding horizon \cite{Misner:book1973}. 
Therefore the approximation of a gravitating compact body as a point-like massive object in curved space-time 
requires the body to be small compared to the radius of curvature of the background space-time \cite{Plyatsko:2011gf,Poisson:2003nc}. In addition, 
the mass must be small enough to ignore its effect on the space-time geometry at large. In the existing literature 
much effort has been put into obtaining effective equations of motion for compact objects by defining a position 
variable which can be interpreted effectively as the material center \cite{Semerak:1999qc,Costa:2014nta}. One then computes the momentum and 
angular momentum in terms of a mass and momentum distribution in a finite neighborhood of this point. 
However, such a position variable is not unique, and moreover it often traces out a complicated world line, 
as shown for example by the well-known helical motion that is a solution of the Mathisson-Papapetrou-Dixon 
equations \cite{Mathisson:1937zz, Costa:2011zn,Costa:2012rr}.

In situations where a point-particle approximation of a spinning and gravitating body is appropriate, a complementary 
approach suggests itself by constructing a lagrangean or hamiltonian mechanics for a mass-point carrying spin 
in a curved space-time. In this letter we have chosen the hamiltonian point of view, as in our opinion this is
most transparent in its results and application. In particular, the closed set of Dirac-Poisson brackets (\ref{2.4}) 
provides a unique and unambiguous starting point for the derivation of equations of motion for any representation
of the spin degrees of freedom, allowing for a large class of physical implementations as fixed by the choice 
of hamiltonian. Two such choices, a minimal and a non-minimal one, have been presented and analyzed in this 
letter. 

The minimal choice of hamiltonian is the one which also describes the geodesic motion of spinless particles. 
With this choice of hamiltonian the spin is covariantly constant along the world line, which is no longer 
geodesic due to spin-orbit coupling. It naturally provides a different implementation of the notion of position 
of the body, one for which now the Pirani vector $Z$ is no longer taken to vanish. The advantage is that in 
terms of this choice of position variable the motion becomes tractable in non-trivial situations of practical 
interest; the motion in Schwarzschild space-time, as analyzed in sect.\ 5, provides a case in point. 
In addition, non-minimal hamiltonians can provide more complicated dynamics, as required for example
for objects with non-vanishing gravimagnetic ratios \cite{Khriplovich:1998ev,Khriplovich:2008ni}. In this case the spin is subject to a kind of gravitational 
Larmor force, making it precess around field-lines of constant curvature. 

The question which effective hamiltonian to use for which physical system now becomes a matter of 
phenomenology. One should either derive the correct effective hamiltonian from first principles, connecting 
the formalism to the specific energy-momentum tensor, or determine it from experiments or observations. 
For the particular case of rotating black holes it could presumably be measured by observing gravitational 
waves from Extreme Mass Ratio binary systems involving a stellar-mass black hole; for a review see \cite{Babak:2014kqa}.  
\vs{3}

\nit
{\bf Acknowledgement} \\
For G.d'A.\ and J.W.v.H.\ this work is part of the research programme {\em Gravitational Physics} of the 
Netherlands Foundation for Fundamental Research on Matter (FOM).
\vs{5}

\bibliographystyle{newutphys}
\bibliography{bibliography_spinletter}

\providecommand{\href}[2]{#2}\begingroup\raggedright\begin{thebibliography}{10}

\bibitem{deSitter:1916zza}
W.~de~Sitter, ``{Einstein's theory of gravitation and its astronomical
  consequences, First Paper},''
{\em Mon.Not.Roy.Astron.Soc.} {\bf 76} (1916)  699--728.

\bibitem{Thomas:1926dy}
L.~Thomas, ``{The motion of a spinning electron},''
\href{http://dx.doi.org/10.1038/117514a0}{{\em Nature} {\bf 117} (1926)  514}.

\bibitem{Frenkel:1926zz}
J.~Frenkel, ``{Die Elektrodynamik des rotierenden Elektrons},''
\href{http://dx.doi.org/10.1007/BF01397099}{{\em Z.Phys.} {\bf 37} (1926)
  243--262}.

\bibitem{Mathisson:1937zz}
M.~Mathisson, ``{Neue mechanik materieller systeme},''
{\em Acta Phys.Polon.} {\bf 6} (1937)  163--2900.

\bibitem{Papapetrou:1951pa}
A.~Papapetrou, ``{Spinning test particles in general relativity. 1.},''
\href{http://dx.doi.org/10.1098/rspa.1951.0200}{{\em Proc.Roy.Soc.Lond.} {\bf
  A209} (1951)  248--258}.

\bibitem{Fock:1954}
V.~Fock and D.~Ivanenko, ``{\"{U}ber eine m\"{o}gliche geometrische Deutung der
  relativistischen Quantentheorie},''
  \href{http://dx.doi.org/10.1007/BF01341739}{{\em Z.Phys.} {\bf 54} (1929)
  798--802}.

\bibitem{Dixon:1970zza}
W.~Dixon, ``{Dynamics of extended bodies in general relativity. I. Momentum and
  angular momentum},''
\href{http://dx.doi.org/10.1098/rspa.1970.0020}{{\em Proc.Roy.Soc.Lond.} {\bf
  A314} (1970)  499--527}.

\bibitem{Wald:1972sz}
R.~M. Wald, ``{Gravitational spin interaction},''
\href{http://dx.doi.org/10.1103/PhysRevD.6.406}{{\em Phys.Rev.} {\bf D6} (1972)
   406--413}.

\bibitem{Hanson:1974qy}
A.~J. Hanson and T.~Regge, ``{The Relativistic Spherical Top},''
\href{http://dx.doi.org/10.1016/0003-4916(74)90046-3}{{\em Annals Phys.} {\bf
  87} (1974)  498}.

\bibitem{Semerak:1999qc}
O.~Semerak, ``{Spinning test particles in a Kerr field. I.},''
\href{http://dx.doi.org/10.1046/j.1365-8711.1999.02754.x}{{\em
  Mon.Not.Roy.Astron.Soc.} {\bf 308} (1999)  863--875}.

\bibitem{Semerak:2007}
K.~Kyrian and O.~Semerak, ``{Spinning test particles in a Kerr field. II.},''
  \href{http://dx.doi.org/10.1111/j.1365-2966.2007.12502.x}{{\em
  Mon.Not.Roy.Astron.Soc.} {\bf 382} (2007)  1922--1932}.

\bibitem{Plyatsko:2011gf}
R.~Plyatsko, O.~Stefanyshyn, and M.~Fenyk, ``{Mathisson-Papapetrou-Dixon
  equations in the Schwarzschild and Kerr backgrounds},''
  \href{http://dx.doi.org/10.1088/0264-9381/28/19/195025}{{\em
  Class.Quant.Grav.} {\bf 28} (2011)  195025},
\href{http://arxiv.org/abs/1110.1967}{{\tt arXiv:1110.1967 [gr-qc]}}.

\bibitem{Schaefer:2004qh}
G.~Schaefer, ``{Gravitomagnetic effects},''
  \href{http://dx.doi.org/10.1023/B:GERG.0000046180.97877.32}{{\em
  Gen.Rel.Grav.} {\bf 36} (2004)  2223},
\href{http://arxiv.org/abs/gr-qc/0407116}{{\tt arXiv:gr-qc/0407116 [gr-qc]}}.

\bibitem{Steinhoff:2010zz}
J.~Steinhoff, ``{Canonical formulation of spin in general relativity},''
  \href{http://dx.doi.org/10.1002/andp.201000178}{{\em Annalen Phys.} {\bf 523}
  (2011)  296--353},
\href{http://arxiv.org/abs/1106.4203}{{\tt arXiv:1106.4203 [gr-qc]}}.

\bibitem{Costa:2012cy}
L.~F.~O. Costa, J.~Nat\'{a}rio, and M.~Zilhao, ``{Spacetime dynamics of
  spinning particles - exact gravito-electromagnetic analogies},''
\href{http://arxiv.org/abs/1207.0470}{{\tt arXiv:1207.0470 [gr-qc]}}.

\bibitem{Bailey:1975fe}
I.~Bailey and W.~Israel, ``{Lagrangian Dynamics of Spinning Particles and
  Polarized Media in General Relativity},''
\href{http://dx.doi.org/10.1007/BF01609434}{{\em Commun.Math.Phys.} {\bf 42}
  (1975)  65--82}.

\bibitem{Barut:book}
A.~O. Barut, {\em {Electrodynamics and classical theory of fields and
  particles}}.
\newblock Mac Millan, New York, 1964.

\bibitem{Duval:1972zza}
C.~Duval, H.~H. Fliche, and J.~M. Souriau {\em C. R. Acad. Sci. Paris.} {\bf
  274} (1972)  1082.

\bibitem{Brink:1976uf}
L.~Brink, P.~Di~Vecchia, and P.~S. Howe, ``{A Lagrangian Formulation of the
  Classical and Quantum Dynamics of Spinning Particles},''
\href{http://dx.doi.org/10.1016/0550-3213(77)90364-9}{{\em Nucl.Phys.} {\bf
  B118} (1977)  76}.

\bibitem{barut:1984}
A.~O. Barut and N.~Zanghi {\em Phys. Rev. Lett..} {\bf 52} (1984)  .

\bibitem{Khriplovich:1996dv}
I.~Khriplovich and A.~Pomeransky, ``{Gravitational interaction of spinning
  bodies, center-of-mass coordinate and radiation of compact binary systems},''
  \href{http://dx.doi.org/10.1016/0375-9601(96)00266-6}{{\em Phys.Lett.} {\bf
  A216} (1996)  7},
\href{http://arxiv.org/abs/gr-qc/9602004}{{\tt arXiv:gr-qc/9602004 [gr-qc]}}.

\bibitem{Khriplovich:1998ev}
I.~Khriplovich and A.~Pomeransky, ``{Equations of motion of spinning
  relativistic particle in external fields},''
  \href{http://dx.doi.org/10.1080/01422419908228843}{{\em Surveys High
  Energ.Phys.} {\bf 14} (1999)  145--173},
\href{http://arxiv.org/abs/gr-qc/9809069}{{\tt arXiv:gr-qc/9809069 [gr-qc]}}.

\bibitem{Barducci:1976qu}
A.~Barducci, R.~Casalbuoni, and L.~Lusanna, ``{Supersymmetries and the
  Pseudoclassical Relativistic electron},''
\href{http://dx.doi.org/10.1007/BF02730291}{{\em Nuovo Cim.} {\bf A35} (1976)
  377}.

\bibitem{Berezin:1976eg}
F.~Berezin and M.~Marinov, ``{Particle Spin Dynamics as the Grassmann Variant
  of Classical Mechanics},''
\href{http://dx.doi.org/10.1016/0003-4916(77)90335-9}{{\em Annals Phys.} {\bf
  104} (1977)  336}.

\bibitem{Salomonson:1978sk}
P.~Salomonson, ``{Supersymmetric Actions for Spinning Particles},''
\href{http://dx.doi.org/10.1103/PhysRevD.18.1868}{{\em Phys.Rev.} {\bf D18}
  (1978)  1868--1880}.

\bibitem{Rietdijk:1992tx}
R.~Rietdijk and J.~van Holten, ``{Spinning particles in Schwarzschild
  space-time},''
\href{http://dx.doi.org/10.1088/0264-9381/10/3/017}{{\em Class.Quant.Grav.}
  {\bf 10} (1993)  575--594}.

\bibitem{Barausse:2009aa}
E.~Barausse, E.~Racine, and A.~Buonanno, ``{Hamiltonian of a spinning
  test-particle in curved spacetime},''
  \href{http://dx.doi.org/10.1103/PhysRevD.85.069904,
  10.1103/PhysRevD.80.104025}{{\em Phys.Rev.} {\bf D80} (2009)  104025},
\href{http://arxiv.org/abs/0907.4745}{{\tt arXiv:0907.4745 [gr-qc]}}.

\bibitem{Khriplovich:1989ed}
I.~Khriplovich, ``{Particle with internal angular momentum in a gravitational
  field},''
{\em Sov.Phys.JETP} {\bf 69} (1989)  217--219.

\bibitem{vanHolten:2006xq}
J.~W. van Holten, ``{Covariant Hamiltonian dynamics},''
  \href{http://dx.doi.org/10.1103/PhysRevD.75.025027}{{\em Phys.Rev.} {\bf D75}
  (2007)  025027},
\href{http://arxiv.org/abs/hep-th/0612216}{{\tt arXiv:hep-th/0612216
  [hep-th]}}.

\bibitem{vanHolten:1990we}
J.~W. van Holten, ``{On the electrodynamics of spinning particles},''
\href{http://dx.doi.org/10.1016/0550-3213(91)90139-O}{{\em Nucl.Phys.} {\bf
  B356} (1991)  3--26}.

\bibitem{Costa:2011zn}
L.~F.~O. Costa, C.~A. Herdeiro, J.~Nat\'{a}rio, and M.~Zilhao, ``{Mathisson's
  helical motions for a spinning particle: Are they unphysical?},''
  \href{http://dx.doi.org/10.1103/PhysRevD.85.024001,
  10.1103/PhysRevD.85.029903}{{\em Phys.Rev.} {\bf D85} (2012)  024001},
\href{http://arxiv.org/abs/1109.1019}{{\tt arXiv:1109.1019 [gr-qc]}}.

\bibitem{Costa:2014nta}
L.~F. Costa and J.~Nat\'{a}rio, ``{Center of mass, spin supplementary
  conditions, and the momentum of spinning particles},''
\href{http://arxiv.org/abs/1410.6443}{{\tt arXiv:1410.6443 [gr-qc]}}.

\bibitem{Pirani:1956tn}
F.~Pirani, ``{On the Physical significance of the Riemann tensor},''
\href{http://dx.doi.org/10.1007/s10714-009-0787-9}{{\em Acta Phys.Polon.} {\bf
  15} (1956)  389--405}.

\bibitem{Khriplovich:2008ni}
I.~Khriplovich, ``{Spinning Relativistic Particles in External Fields},''
\href{http://arxiv.org/abs/0801.1881}{{\tt arXiv:0801.1881 [gr-qc]}}.

\bibitem{vanHolten:1992hs}
J.~W. van Holten, ``{Relativistic dynamics of spin in strong external
  fields},''
\href{http://arxiv.org/abs/hep-th/9303124}{{\tt arXiv:hep-th/9303124
  [hep-th]}}.

\bibitem{Rudiger:1981}
{R\"{u}diger, R.}, ``{Conserved quantities of spinning test particles in
  General Relativity I},'' {\em Proc.Roy.Soc. Series A} {\bf 375} (1981)
  no.~1761, 185--193.

\bibitem{Rudiger:1983}
{R\"{u}diger R.}, ``{Conserved quantities of spinning test particles in General
  Relativity II},'' {\em Proc.Roy.Soc. Series A} {\bf 385} (1983) no.~1788,
  229--239.

\bibitem{Gibbons:1993ap}
G.~Gibbons, R.~Rietdijk, and J.~van Holten, ``{SUSY in the sky},''
  \href{http://dx.doi.org/10.1016/0550-3213(93)90472-2}{{\em Nucl.Phys.} {\bf
  B404} (1993)  42--64},
\href{http://arxiv.org/abs/hep-th/9303112}{{\tt arXiv:hep-th/9303112
  [hep-th]}}.

\bibitem{VanHolten:1997ur}
J.~W. Van~Holten, ``{Gravitational waves and black holes: An Introduction to
  general relativity},'' {\em Fortsch.Phys.} {\bf 45} (1997)  439--516,
\href{http://arxiv.org/abs/gr-qc/9704043}{{\tt arXiv:gr-qc/9704043 [gr-qc]}}.

\bibitem{Bini:2005nt}
D.~Bini, F.~de~Felice, A.~Geralico, and R.~T. Jantzen, ``{Spin precession in
  the Schwarzschild spacetime: Circular orbits},''
  \href{http://dx.doi.org/10.1088/0264-9381/22/14/007}{{\em Class.Quant.Grav.}
  {\bf 22} (2005)  2947--2970},
\href{http://arxiv.org/abs/gr-qc/0506017}{{\tt arXiv:gr-qc/0506017 [gr-qc]}}.

\bibitem{Bini:2014poa}
D.~Bini, A.~Geralico, and R.~T. Jantzen, ``{Spin-geodesic deviations in the
  Schwarzschild spacetime},''
  \href{http://dx.doi.org/10.1007/s10714-010-1111-4}{{\em Gen.Rel.Grav.} {\bf
  43} (2011)  959},
\href{http://arxiv.org/abs/1408.4946}{{\tt arXiv:1408.4946 [gr-qc]}}.

\bibitem{d'Ambrosi:preparation}
G.~d'Ambrosi, S.~Satish~Kumar, and J.~W. van Holten {\em in preparation} (2015)
   .

\bibitem{Khriplovich:1997ni}
I.~Khriplovich and A.~Pomeransky, ``{Equations of motion of spinning
  relativistic particle in external fields},''
  \href{http://dx.doi.org/10.1134/1.558554}{{\em J.Exp.Theor.Phys.} {\bf 86}
  (1998)  839--849},
\href{http://arxiv.org/abs/gr-qc/9710098}{{\tt arXiv:gr-qc/9710098 [gr-qc]}}.

\bibitem{Heinemann:1996aq}
K.~Heinemann, ``{On Stern-Gerlach forces allowed by special relativity and the
  special case of the classical spinning particle of Derbenev-Kondratenko},''
\href{http://arxiv.org/abs/physics/9611001}{{\tt arXiv:physics/9611001
  [physics]}}.

\bibitem{Misner:book1973}
C.~W. {Misner}, K.~S. {Thorne}, and J.~A. {Wheeler}, {\em {Gravitation}}.
\newblock 1973.

\bibitem{Poisson:2003nc}
E.~Poisson, ``{The Motion of point particles in curved space-time},'' {\em
  Living Rev.Rel.} {\bf 7} (2004)  6,
\href{http://arxiv.org/abs/gr-qc/0306052}{{\tt arXiv:gr-qc/0306052 [gr-qc]}}.

\bibitem{Costa:2012rr}
L.~F.~O. Costa, J.~Nat\'{a}rio, and M.~Zilhao, ``{Mathisson's helical motions
  demystified},'' \href{http://dx.doi.org/10.1063/1.4734436}{{\em AIP
  Conf.Proc.} {\bf 1458} (2011)  367--370},
\href{http://arxiv.org/abs/1206.7093}{{\tt arXiv:1206.7093 [gr-qc]}}.

\bibitem{Babak:2014kqa}
S.~Babak, J.~R. Gair, and R.~H. Cole, ``{Extreme mass ratio inspirals:
  perspectives for their detection},''
\href{http://arxiv.org/abs/1411.5253}{{\tt arXiv:1411.5253 [gr-qc]}}.

\end{thebibliography}\endgroup

\end{document}